\documentclass[
    ,final            
  ]
  {aipproc}

\layoutstyle{6x9}

\usepackage{mathbbol}
\usepackage{epsfig}
\usepackage{color}
\usepackage{amsfonts}
\usepackage{amsmath}
\usepackage{amssymb}
\def\lsi{\raise0.3ex\hbox{$<$\kern-0.75em\raise-1.1ex\hbox{$\sim$}}}
\def\gsi{\raise0.3ex\hbox{$>$\kern-0.75em\raise-1.1ex\hbox{$\sim$}}}

\def\xlf{\raisebox{+0.2em}{\boldmath{$\chi$}}\hspace{-0.2ex}\raisebox{-0.2em}{L}
\hspace{-1.5ex}\raisebox{+0.14em}{F}\hspace{2mm}} 

\begin{document}

\title{Lattice Gauge Actions for Fixed Topology}

\author{\xlf Collaboration: W.~Bietenholz}{
  address={Institut f\"ur Physik, Humboldt 
Universit\"at zu Berlin, Newtonstr.\ 15, 12489 Berlin, Germany}
}

\author{K.~Jansen}{
  address={NIC/DESY Zeuthen, Platanenallee 6, D-15738 Zeuthen, Germany}
}

\author{K.-I.~Nagai}{
  address={NIC/DESY Zeuthen, Platanenallee 6, D-15738 Zeuthen, Germany}
}

\author{S.~Necco}{
  address={Centre de Physique Th\'{e}orique, Luminy, Case 907, F-13288 Marseille Cedex 9, France}
}

\author{L.~Scorzato}{
  address={Institut f\"ur Physik, Humboldt 
Universit\"at zu Berlin, Newtonstr.\ 15, 12489 Berlin, Germany}
}

\author{S.~Shcheredin}{
  address={Institut f\"ur Physik, Humboldt 
Universit\"at zu Berlin, Newtonstr.\ 15, 12489 Berlin, Germany}
}

\begin{abstract}
We test a set of lattice gauge actions for QCD that suppress small plaquette values
and in this way also suppress transitions between topological sectors.
This is well suited for simulations in the 
$\epsilon$-regime and it is expected to help in
numerical simulations with dynamical quarks. 
\end{abstract}

\maketitle

Our aim is to study the possibility of simulating lattice QCD
with a gauge action that strongly reduces the occurrence of
small plaquette values.
A gauge background with such a feature is expected to improve 
the locality properties \cite{Hernandez:1998et} of the Overlap-Dirac operator $D_{\rm ov}$ 
\cite{Neuberger:1997fp}.
By the same argument one also expects to ease 
the  numerical evaluation of $D_{\rm ov}$ itself, and help 
in general dynamical simulations.
It can be proven 
\cite{Hernandez:1998et,Neuberger:1999pz} that 
as long as all plaquette values in a gauge configuration satisfy:
$ S_{P} := 1 - \frac{1}{3} {\rm Re ~ Tr} (U_{P}) <  1/ 20.5 $
then no change of the topological sector is possible.
Hence a suppression of low plaquette values entails a suppression of
$Q_{\rm top}$ changes.

Simulations constrained  in a fixed topological sector can 
be problematic for evaluating physical observables in  QCD, 
where all sectors have to be
taken into  account with the correct weight. However such a
constraint is perfectly suited for studying QCD in the 
$\epsilon$-regime \cite{Gasser:1987ah}, 
where predictions
exist for observables defined in fixed topological
sectors \cite{Leutwyler:1992yt}, which turns the limitation
mentioned above into an advantage.
However there are some conditions.
The physical volume should be at least
$L \gtrsim 1.1 ~ {\rm fm}$ \cite{Bietenholz:2003mi,Bietenholz:2003bj}. 
Moreover, in order to reach very
small pion masses, the chiral properties of the Dirac operator
are crucial. Finally a sound definition of 
$Q_{\rm top} $  is important to
compare with predictions in fixed topological sectors.
All of these requirements are provided by Ginsparg-Wilson fermions.
They have an exact, lattice modified chiral symmetry \cite{Luscher:1998pq}, and the
fermionic index defines  $ Q_{\rm top} $ \cite{Hasenfratz:1998ri}.
Results in the $\epsilon-$regime with Ginsparg-Wilson fermions
were obtained for the Dirac spectrum 
\cite{Bietenholz:2003mi,Giusti:2003gf,Galletly:2003vf}
and for meson correlation functions 
\cite{Bietenholz:2003bj,Giusti:2003iq,Giusti:2004yp,Giusti:2004bf}, 
which were compared with quenched Chiral Perturbation Theory 
\cite{Damgaard:2001js,Damgaard:2002qe,Colangelo:2004sc}.

Simple examples of gauge actions that suppress small
plaquette values (still expected to be in the same universality class as $S_P$) are
\begin{eqnarray}
\beta ~ S^{\mbox{\tiny hyp}}_{\varepsilon,n} (U_P) &=& 
\beta ~ \frac{S_{P}}{(1 - \varepsilon^{-1} S_{P})^{n}} 
\hspace{1cm}
\mbox{if }\, S_P < \varepsilon \mbox{, and } +\infty \mbox{ otherwise} \label{Shyp}\\
\beta ~ S^{\mbox{\tiny pow}}_{\varepsilon,n}(U_P) &=& 
\beta ~ S_P  + \varepsilon^{-1}  S_P^n  \label{Spow}\\
\beta ~ S^{\mbox{\tiny exp}}_{\varepsilon,n}(U_P) &=& 
\beta ~ S_P  \exp{[ \varepsilon^{-1} S_P^n ]} \label{Sexp}
\end{eqnarray}
The first choice above (for $n =1$)
was introduced by M.\ L\"{u}scher for conceptual purposes 
\cite{Luscher:1998du}, and applied by Fukaya and Onogi in Schwinger model 
simulations \cite{Fukaya:2003ph,Fukaya:2004kp}. 
The question is whether one can conciliate the advantages mentioned above, 
with reasonable lattice sizes (say $La \sim 1-2 ~ {\rm fm}$), 
without increasing lattice artifacts and 
with a correct and  reasonably decorrelated sampling of interesting observables.
A first report of our ongoing study was presented in Ref. \cite{Shcheredin:2004xa}.

{\em Results.}
Since gauge actions of the type (\ref{Shyp},\ref{Spow},\ref{Sexp}) are non-linear in the
link variables, the heat-bath algorithm cannot be applied.
Instead we use a local HMC algorithm \cite{Marenzoni:1993im}, which is competitive with
heat-bath in the standard case. HMC trajectories have discretization 
$dt$ quoted in Table \ref{r0Tab}, and the trajectory length is 1.
The volume is chosen to $16^4$ in order to allow a reliable
determination of $r_0/a \simeq 7$ and in order have mild finite volume effects.
We computed $r_0/a$ following a standard procedure \cite{Necco:2001gh}.
Our preliminary results are summarized in Table \ref{r0Tab}.
\begin{table}
\begin{tabular}{|c|c|c|c|c|c|c|c|c|c|}
\hline
{\small $\varepsilon^{-1}$}   & {\small $\beta$}  & {\small $r_{0}/a$} & {\small $\beta_{W}$ } & 
{\small $\tau^{\rm plaq}$} & {\small $\tau^{\rm plaq}(\beta_W)$} & {\small $f_J$} & {$dt$} & {\small Acceptance}\\
\hline
\hline
{\small 0   } & {\small 6.18} & {\small 7.14(3)} & {\small 6.18} & {\small 7(1)}& {\small 7(1)} & {\small  {\em 0.015}} & {\small 0.1 } & {\small $> 99 \%$} \\
\hline
{\small 1.00} & {\small 1.5} & {\small 6.6(2)} & {\small 6.13(2)} & {\small 2.0(1)}& {\small n.a.} & {\small 0.0027} & {\small 0.05} & {\small $> 99 \%$} \\
\hline
{\small 1.18} & {\small 1.0} & {\small 7.2(2)} & {\small 6.18(2)} & {\small 1.3(1)}& {\small 7(1)} & {\small 0.0014} & {\small 0.02 - 0.01} & {\small $> 99 \%$} \\
\hline
{\small 1.25} & {\small 0.8} & {\small 7.0(1)} & {\small 6.17(1)} & {\small 1.1(1)} & {\small 9(1)} & {\small 0.0025} & {\small 0.1} & {\small $> 99 \%$} \\
\hline
{\small 1.52} & {\small 0.3} & {\small 7.3(4)} & {\small 6.19(4)} & {\small 0.8(1)} & {\small 7(1)} & {\small 0.0008} & {\small 0.1} & {\small $\sim 95 \%$} \\
\hline
{\small 1.64} & {\small 0.1} & {\small 6.8(3)} & {\small 6.15(3)} & {\small 1.0(1)} & {\small n.a.} & {\small 0.0007} & {\small 0.1} & {\small $\sim 65 \%$} \\
\hline
\end{tabular}
\caption{{\it Results for the $S^{\mbox{\tiny hyp}}_{\varepsilon,1}$ on a $16^{4}$ lattice
for various $\varepsilon^{-1}$ and $\beta$. $\beta_W$ is the coupling resulting in the same $r_0$ with the Wilson action.
}}
\label{r0Tab}
\end{table}
We estimated the topological charge with cooling and searching for the first
plateau \cite{Ilgenfritz:1985dz}.  Since we cannot reliably measure
the autocorrelation of the $Q_{\rm top}$, we quote -- as an
indicator of stability -- the number of jumps of $Q_{\rm top}$
divided by the number of trajectories in the full history ($f_J$).
Since we save a configuration only every 50 trajectories, the measured 
$f_J$ is only a lower bound on the frequency of jumps, which can be reliable
only for $f_J \ll 0.01$ (which is the interesting case for us). 
In particular for the Wilson action at $\beta \sim 6.17$ we expect 
$f_J \gg 0.015$, which we measure.
The stability of $Q_{\rm top}$  has to be compared with the
autocorrelation of a typical observable (we quote the plaquette value
under $\tau^{\rm plaq}$). It is interesting to see that the latter is strongly
decreased for non-zero $\varepsilon^{-1}$ (at fixed $r_0/a$).
We have also studied actions of type (\ref{Sexp}), which have a smooth 
bound on the plaquette value,
and therefore are better suited for efficient simulations with global HMC and
dynamical fermions. Results will be presented elsewhere.
We also checked that the results are consistent independently 
on the starting configuration. This is important because
the constraint on the plaquette value could in principle generate
more obstructions than the topological ones, and this would not
be noticed simply from the autocorrelation.

It has been  pointed out \cite{Creutz:2004ir} that the actions
(\ref{Shyp}) do not allow for the existence of a positive definite
transfer matrix. However we checked that all the actions
(\ref{Shyp},\ref{Spow},\ref{Sexp}) have {\em site}-reflection positivity,
which at least ensures the existence of a positive squared transfer matrix 
\cite{Montvay:1994cy}.
Moreover we have checked that the
unphysical behaviour of the short distance force -- which was observed
in some cases, and related to the lack of a positive
transfer matrix \cite{Necco:2003vh} -- does not appear in our cases.

We stress that -- inspired by Ref. \cite{Hernandez:1998et} and by simplicity --
our search is restricted to actions which are functions of the plaquette value,
thus excluding other alternatives, which also affect the stability
of $Q_{\rm top}$, as for example in Ref.
\cite{Orginos:2001xa,Aoki:2002vt}.

{\em Conclusions.}
Topology conserving gauge actions
could be highly profitable in QCD simulations.
The suppression of small plaquette values may speed
up the simulations with dynamical quarks.
A stable $Q_{\rm top}$ is useful
in particular in the $\epsilon$-regime.
We are investigating such actions,
in view of the physical scale and the topological stability.

{\em Ackowledgements}. 
We thank M.~Creutz, H.~Fukaya, S.~Hashimoto, M.~L{\"u}scher, T.~Onogi and  R.~Sommer 
for helpful discussions and correspondence. 
The Deutsche Forschungsgemeinschaft through SFB-TR 9-03,
and the European Union under contract HPRN-CT-2002-00311 (EURIDICE),
are acknowledged for financial support.

\bibliographystyle{h-physrev4}

\bibliography{topocons}

\IfFileExists{\jobname.bbl}{}
 {\typeout{}
  \typeout{******************************************}
  \typeout{** Please run "bibtex \jobname" to optain}
  \typeout{** the bibliography and then re-run LaTeX}
  \typeout{** twice to fix the references!}
  \typeout{******************************************}
  \typeout{}
 }

\end{document}